\documentclass[12pt]{article}

\usepackage{amsmath}
\usepackage{amssymb}
\usepackage{mathrsfs}
\usepackage{epsfig}
\usepackage{latexsym}
\usepackage{amssymb}
\usepackage{amsmath}
\usepackage{epsfig}
\usepackage{natbib}
\usepackage[dvips]{color}

\newcommand{\p}{\ensuremath{\mathbb {{}^\prime}}}

\def\half{\frac{1}{2}}

\begin{document}

\begin{titlepage}

\setcounter{page}{1} \baselineskip=15.5pt \thispagestyle{empty}

\begin{flushright}
ULB-TH/07-17\\ %hep-th/yymmnnn
arXiv:0704.3484 [hep-th]
\end{flushright}
\vfil

\begin{center}
{\LARGE Cascading Quivers from Decaying D-branes}
\end{center}
\bigskip\

\begin{center}

{\large Jarah Evslin\footnote{ jevslin@ulb.ac.be}, Chethan Krishnan\footnote{ Chethan.Krishnan@ulb.ac.be} 
and Stanislav Kuperstein\footnote{skuperst@ulb.ac.be}}

\end{center}

\begin{center}
\textit{{\it Physique Th\'eorique et Math\'ematique,
International Solvay
Institutes, \\ Universit\'e Libre de Bruxelles, ULB Campus Plaine C.P. 
231, 
B--1050 Bruxelles,
Belgium}}
\end{center} \vfil

\noindent 
We use an argument analogous to that of Kachru, Pearson and Verlinde to 
argue that cascades in $L^{a,b,c}$ quiver gauge theories 
always preserve the form of the quiver, and that all gauge groups drop at 
each step by the number $M$ of fractional branes.
In particular, we demonstrate that an NS5-brane that sweeps out the $S^3$ 
of the base of $L^{a,b,c}$ destroys $M$ D3-branes.

\vfil

\end{titlepage}

%---------------------------------------------------------------------------------------------------
\tableofcontents

\pagestyle{headings}

%---------------------------------------------------------

\section{\bf Introduction}

Klebanov and his collaborators have demonstrated \cite{KW,KN,KT,KS}
that type IIB string theory on $AdS_5\times T^{1,1}$ is 
holographically dual to a cascading gauge theory. 
Three years later it was conjectured that there 
are more cascading gauge theories dual to the product of $AdS_5$ and an infinite family of 
Sasaki-Einstein manifolds $L^{a,b,c}$ that generalise $T^{1,1}$
\cite{Gauntlett:2004yd,Bertolini:2004xf,Benvenuti:2004dy,Cvetic:2005ft,
       Benvenuti:2005cz,Benvenuti:2005ja,Franco:2005sm,Cvetic:2005vk,
       Butti:2005sw,Martelli:2005wy,Berenstein:2005xa,Franco:2005zu,Bertolini:2005di,Brini:2006ej}. 
However these cases are complicated 
by the fact that multiple gauge couplings become strong simultaneously, 
and so one does not quite know how to perform the duality, for example it may be 
that for a fixed $L^{a,b,c}$ there exist a network of walls separating domains 
of initial values of gauge couplings which exhibit different cascades \cite{Fiol:2002ah,Franco:2003ja,Franco:2004jz}.  
We will argue that only one such pattern of cascades appears to be consistent with RR charge 
conservation in the dual gravity description in the case of compactifications on $AdS_5\times L^{a,b,c}$ 
with $a$, $b$, $c$
and $d \equiv a+b-c$ 
are relatively prime and also in the cases with $c=d$, in which $L^{a,b,c}$ is a $Y^{p,q}$ \cite{Cvetic:2005vk}.

The manifolds $L^{a,b,c}$ are similar to $T^{1,1}$, the base of the conifold.  
Topologically they are identical, if $a$, $b$, $c$ and $d$ are relatively 
prime then $L^{a,b,c}$ 
is diffeomorphic to $T^{1,1}$ 
and both are diffeomorphic to $S^2\times S^3$ \cite{CdO,KW}.  An explicit diffeomorphism relating $T^{1,1}$ 
to $S^2\times S^3$ was presented in \cite{EK}.  However the metrics on $T^{1,1}$ and $L^{a,b,c}$ 
are not equivalent and as a result the world-volume gauge symmetries of the gauge theories dual 
to the $AdS_5\times T^{1,1}$ and $AdS_5\times L^{a,b,c}$ backgrounds are 
very different.  
The gauge theories dual to $L^{a,b,c}$ compactifications are far more complicated and 
their vacuum structures are not understood, 
which is an obstruction to the analysis of their cascades of Seiberg dualities. 

While it remains quite difficult to determine the vacuum structure of these theories, 
we will argue that the topology (in fact just the homology) of $L^{a,b,c}$ 
along with the fluxes present 
in the compactification already places a strong constraint on the dualities allowed in the dual gauge theory.  
This constraint arises by imposing that the dualities arise from processes that conserve RR charge, 
generalising the NS5-brane nucleation in the $T^{1,1}$ case which was presented in \cite{KPV,MeCascade,Evslin:2007ti}.  
More specifically, consider $T^{1,1}$, which we recall again is diffeomorphic 
to $S^2\times S^3$, with $M$ units of RR 3-form flux $F_3$

\begin{equation}
\int_{S^3} F_3=M. \label{eq:g3}
\end{equation}
The dual gauge theory, which intuitively lives on $N$ D3-branes that are 
put at points on $L^{a,b,c}$, 
has a $SU(N)\times SU(N+M)$ gauge symmetry.  
Now consider an NS5-brane that wraps the {4-dimensional} horizon and also wraps a contractible 
2-sphere at fixed latitude $\theta$ in the $S^3$ of $T^{1,1}$.  
There is a family of such configurations, parameterised by the latitude $\theta$.  
The central result of ~\cite{KPV,MeCascade} is that the parameter $\theta$ 
parameterises the re-normalisation group direction, 
and in particular as the gauge theory flows into the IR the 2-sphere 
nucleates at the south pole of the $S^3$, moves up to the equator and then shrinks down 
to nothing again at the north pole.  
We will refer to this process as a MMS instanton, as it was first described in \cite{MMS}.  
A somewhat simplified version of this system was analysed classically in \cite{BDS}.

When the NS5-brane shrinks down to nothing there are $M$ (anti) D3-branes left.  
One can see this immediately using RR charge conservation.  
The NS5-brane sources $H$ flux, and by Gauss' law the MMS instanton 
increases the $H$ flux by one Dirac unit.  
The total RR 3-charge is equal to the sum of the brane contribution, equal to the number of D3-branes, 
plus a bulk contribution
\begin{eqnarray}
Q_{RR}=N_{D3}+\int H\wedge F_3.
\end{eqnarray}
When $H$ increases by a single unit, 
this wedge product increases by $M$ units, as one finds for example by using Poincar\'e duality 
to express the integral of $\Delta H\wedge F_3$ as the integral (\ref{eq:g3}) of $F_3$ over the $S^3$ 
swept out by the NS5.  The total RR 3-charge must be conserved, 
and so if the bulk charge increases by $M$ units then the brane charge must decrease by $M$ units, 
meaning that there are $M$ less D3-branes, leaving $N-M$.  The new gauge 
group 
is then $SU(N-M)\times SU(N-M+M)=SU(N-M)\times SU(N)$.  
Alternately one may use the NS5-brane worldvolume theory to see that the instanton leaves $M$ anti-branes.  
The worldvolume Wess-Zumino term
\begin{equation}
S_{\text {NS5}}\supset\int C_2\wedge C_4
\end{equation}
implies that $C_2$ is an electric source for the RR 4-form connection $C_4$, in other words, 
it carries D3-brane charge.  Using (\ref{eq:g3}) and Stoke's theorem one finds that the integral of 
$C_2$ over the 2-sphere wrapped by the NS5-brane decreases by $M$ units during the MMS instanton, 
and so the D3-brane charge decreases by $M$ units, leaving $M$ anti-branes when the NS5 finally collapses.

In the $T^{1,1}$ case, only one simple gauge group is strongly 
coupled in the IR, 
which allows one to treat the other as a global symmetry and so find the 
Seiberg duality directly 
in the field theory, choosing the root of the baryonic branch in the sense 
of \cite{APS} and thus 
demonstrating that the duality is not only allowed by 
RR charge conservation but actually provides a weakly coupled description of 
the strongly coupled IR gauge theory.  
In general it is not certain that an allowed transition provides another description, 
and even if it does then there is no reason to believe that such a description should always be weakly coupled.
We will not be so ambitious.

The MMS instanton is easily generalised to $L^{a,b,c}$, without recourse to the details 
of the $L^{a,b,c}$ gauge theory.  One need only know that  $L^{a,b,c}$ is diffeomorphic 
to $S^2\times S^3$ and that (\ref{eq:g3}) still holds, which is a consequence of 
the fact that the D5-branes wrap the $S^2$ which has intersection number one with the $S^3$. 
The above argument then implies that D3-brane charge is preserved modulo $M$, 
and so cascades are allowed which change the number $N$ of D3-branes by an integral multiple of $M$.  
In particular this only leaves room for a single family of cascades, 
and so appears to exclude duality walls for example.  
We will see that the allowed cascades all preserve the form of the corresponding quiver
\footnote{This ties in well with the expectations from the dimer models \cite{Hanany:2005ve,Franco:2005sm}.}.  
One may object that these cascades only describe baryonic vacua, 
and that RR charge must also be preserved in the dual descriptions of the 
mesonic vacua which correspond to distinct dual theories, 
however these vacua are dual to topologically distinct compactifications and so escape. 

In Sec.~\ref{revsec} we review Seiberg dualities in cascading quivers
by considering the specific example of the $Y^{2,1}$ gauge theory dual.  
Then in Sec.~\ref{calcsec} we will demonstrate that there exists a 3-form on $L^{a,b,c}$ 
that satisfies (\ref{eq:g3}) and is $(2,1)$, 
as is required by supersymmetry \cite{Grana:2000jj,Grana:2001xn}.  
We conclude in Sec.~\ref{conc} with comments on the generalisation to compactifications 
on other spaces and D7-brane processes, and possible resolutions to the apparent contradiction 
with the duality wall literature.

%---------------------------------------------------------

\section{\bf Seiberg Duality in Quiver Gauge Theories} \label{revsec}

The superconformal quiver gauge theories dual to $Y^{p,q}$ spaces were 
first constructed in \cite{Benvenuti:2004dy} generalising the specific 
case of $Y^{2,1}$ \cite{Martelli:2004wu, Feng:2000mi}, and this 
was later further generalised to include the dual quivers for 
all $L^{a,b,c}$ spaces \cite{Benvenuti:2005ja, 
Franco:2005sm, Butti:2005sw}. Cascading Seiberg dualities in these field 
theories were discussed 
by many authors, and the supergravity 
duals of these cascades have also been considered \cite{HEK, 
Burrington:2005zd, Martelli:2005wy, Gepner:2005zt}.

The discussion of cascades in these gauge theories is 
more involved than in the familiar case of the conifold, because in the 
latter the quiver diagram has only two nodes. One of 
these two nodes is strongly coupled in the IR (the one-loop beta function 
is positive), while the other one is weakly coupled. So 
the choice of the node on which one should Seiberg dualize is clear 
in the gauge theory picture. Indeed, after the duality, 
we end up with the same node structure that we started with, but with 
shifted ranks for the gauge groups, and the 
process continues all the way to the base of the cascade, where we lose 
one of the nodes (at least when $N$ is a multiple of $M$) and the cascading comes to an end, resulting in chiral 
symmetry breaking \cite{KS}. This ties in well with the 
picture presented by the Klebanov-Strassler supergravity solution dual to 
the gauge theory: there, one finds a radial dependence of the 5-form flux, 
which results in a logarithmic running of the effective number of 
D3-branes.

But in the case of the $L^{a,b,c}$ quivers, the situation is much less 
clear, and the choice of the cascade step could depend, in 
principle, on which node one chooses to dualize on. We will explain 
this in more detail by using the specific example of the $Y^{2,1}$ quiver  
in the reminder of this section. One purpose of 
this paper is to present a dual geometrical argument that gives us a 
natural way to choose the ``right'' cascade, dual to the supergravity 
description. 

\begin{figure}
\begin{center}
\includegraphics{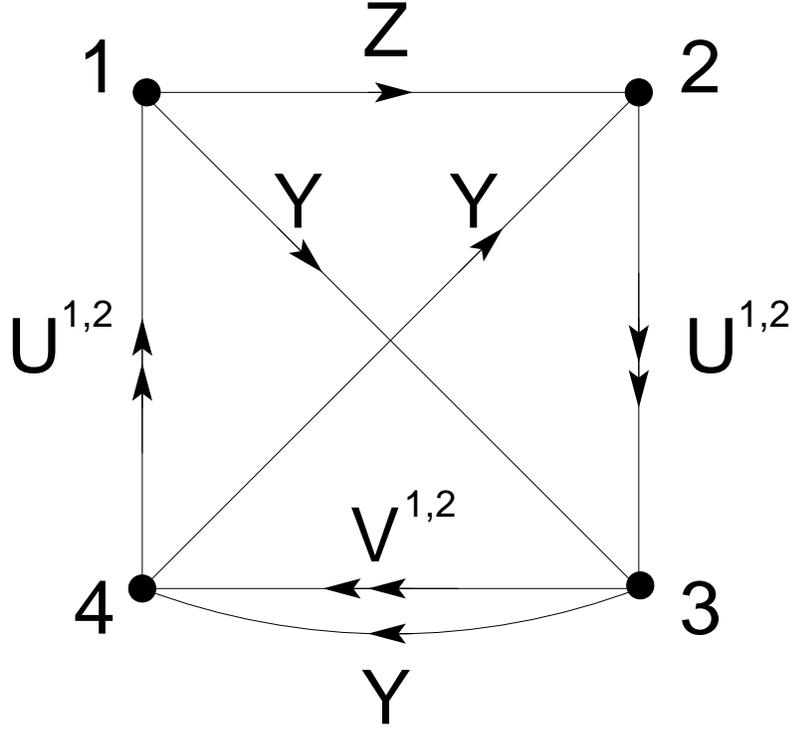}
\caption{The Quiver Diagram for $Y^{2,1}$ theory}
\label{Y21quiver}
\end{center}
\end{figure}

Let us now turn to the explicit example of $Y^{2,1}$. The quiver diagram 
for the theory is shown in figure \ref{Y21quiver}. The diagram 
represents an $\mathcal N=1$ gauge theory, where each node 
corresponds to a gauge group, and each arrow is a chiral bi-fundamental 
superfield, denoted by $U^\alpha, V^{\alpha}, Y$ and $Z$ in the figure, 
$\alpha=1, 2$. 
If the gravity description corresponds to the simplest case, namely that 
of $N$ D3-branes probing the apex of the cone over $Y^{2,1}$, then the 
worldvolume theory on the D3-branes is superconformal, and all the gauge 
groups are $SU(N)$. To trigger the RG-flow that results in the cascade, we 
add $M$ D5-branes and break the conformal invariance. It turns out that 
this changes the 
gauge groups to 
\begin{eqnarray}
SU(N)_1\times SU(N+M)_3\times SU(N+2M)_4\times SU(N+3M)_2,
\end{eqnarray}
where the subscripts serve as a 
book-keeping device to keep track of the node associated to 
the corresponding gauge group in our quiver diagram. The superpotential 
for the theory is the sum of all the gauge invariant cubic and 
quartic operators in the fields listed above. For the case at hand, 

\begin{equation}
\label{eq:Y21superpotential}
W\sim\epsilon_{\alpha\beta}U^{\alpha}_{41}Y_{13}
V^{\beta}_{34}
+\epsilon_{\alpha\beta}V^{\alpha}_{34}Y_{42}U^{\beta}_{23}
+\epsilon_{\alpha\beta}U^{\alpha}_{23}Y_{34}U^{\beta}_{41}Z_{12}.
\end{equation}
The one-loop NSVZ beta function for the running gauge couplings 
for the various nodes can be computed from the usual formula,

\begin{equation}
\label{eq:NSVZ}
\beta_i\equiv\frac{d(8\pi^2/g_i^2)}{d\log 
\mu}=\frac{3T(G)-\sum_iT(r_i)(1-2\gamma_i)}{1-
\frac{g_i^2}{8\pi^2}T(G)}.
\end{equation}
Following Klebanov-Strassler and ignoring the denominator, and using the 
relation $\gamma_i=\frac{3}{2}R_i-1$ relating anomalous dimensions and 
R-charges, we find the following beta functions for the various nodes:
\begin{eqnarray}
\label{eq:beta}
\beta_1&=&3M+\frac{3M}{2}\big[6(R_U-1)+2(R_Y-1)+4(R_Z-1)\big] \nonumber\\
\beta_2&=&12M+\frac{3M}{2}\big[4(R_U-1)+3(R_Y-1)+(R_Z-1)\big] \nonumber \\
\beta_3&=&6M+\frac{3M}{2}\big[6(R_V-1)+8(R_U-1)+4(R_Y-1)\big] \nonumber \\
\beta_4&=&9M+\frac{3M}{2}\big[4(R_V-1)+2(R_U-1)+6(R_Y-1)\big]. \nonumber
\end{eqnarray}
For each node, the gauge groups on the other nodes act as 
effective flavours. In the calculation, we have used the fact that 
$R_U+R_V+R_Y=2$ and 
$2R_U+R_Y+R_Z=2$, which are determined from the conditions for 
conformality when there are no D5-branes.

We can look up the R-charges of the various fields in 
\cite{Bertolini:2004xf}, and the result is (for the specific case of 
$Y^{2,1}$):

\begin{equation}
\label{eq:R-charges}
R_Y=\frac{(-9+3\sqrt{13})}{3}, \, 
R_Z=\frac{(-17+5\sqrt{13})}{3}, \ R_U=\frac{4(4-\sqrt{13})}{3}, \ 
R_V=\frac{(-1+\sqrt{13})}{3}.
\end{equation}
From these explicit values, it follows immediately that nodes 2 and 4 
are both strongly coupled in the infrared, unlike the case of the 
conifold where there was only one gauge coupling that blew up as we 
flowed down along the RG flow. So here the choice of the node to  
Seiberg dualize is in general initial condition dependent. But as 
observed in \cite{HEK}, if we choose to dualize on the node with 
the largest number of colours (in our case, this would be 
node 2), we end up with a quiver that is self-similar to the original one, 
and the usual logic of the cascade still goes through.  This choice has a natural interpretation in 
terms of the D-brane decays that are allowed by K-theory in the dual geometry, it relates D-branes wrapping distinct 
homology classes that represent the same twisted K-theory class, as in 
~\cite{MeCascade,kreview}.

It is also important to note that with this choice of the node, the form 
of the superpotential is also unchanged after Seiberg duality, as we 
will now quickly demonstrate. To Seiberg dualize around node 2, we 
introduce 
meson fields $M^{\alpha}_{43}\equiv Y_{42}U^\alpha_{23}$ and $N^{\alpha}_{13}\equiv Z_{12}U^\alpha_{23}$ 
corresponding to 
the branches 
1-2-3 and 4-2-3 that pass through node 2, and dual quarks ${\tilde 
q_{24}}, {\tilde q_{21}}, q^\beta_{32}$ 
corresponding to the legs that start at node 2. The superpotential for 
the dual theory with these fields will have the pieces 
(\ref{eq:Y21superpotential}) written in terms of the new fields, plus the 
pieces that couple the mesons and the dual quarks as dictated by 
the recipe for Seiberg duality:

\begin{eqnarray}
\label{eq:Wtemp}
W_{{\rm temp}}\sim\epsilon_{\alpha\beta}U^{\alpha}_{41}
Y_{13}V^{\beta}_{34}+\epsilon_{\alpha\beta}V^{\alpha}_{34}M^{\beta}_{43}
+\epsilon_{\alpha\beta}U^{\alpha}_{41}N^{\beta}_{13}Y_{34}
+\epsilon_{\alpha\beta}{\tilde q_{24}}M^{\alpha}_{43} q^\beta_{32}
+\epsilon_{\alpha\beta}{\tilde q_{21}}N^{\alpha}_{13} q^\beta_{32}.
\end{eqnarray}
The $VM$-term is a mass term, and since we are after the IR physics, we 
integrate it out by setting 
\begin{eqnarray}
\frac{\partial W}{\partial V^{\alpha}_{34}}=0 &\implies& 
U^{\alpha}_{41}Y_{13}=M^{\alpha}_{43}, \nonumber \\
\frac{\partial W}{\partial M^{\beta}_{43}}=0 &\implies&
V^{\alpha}_{34}=q^{\alpha}_{32}{\tilde q_{24}}. \nonumber
\end{eqnarray}
The superpotential now looks like

\begin{equation}
\label{eq:Wnew}
W_{{\rm new}}\sim\epsilon_{\alpha\beta}U^{\alpha}_{41}N^{\beta}_{13}Y_{34}
+\epsilon_{\alpha\beta}{\tilde q_{21}}N^{\alpha}_{13}q^\beta_{32}
+\epsilon_{\alpha\beta}U^{\alpha}_{41}Y_{13}q^\beta_{32}
{\tilde q_{24}}, 
\end{equation}
which (after some identifications) is of the same form as 
(\ref{eq:Y21superpotential}).

The crucial thing to notice is that this works only if we choose to 
dualize on node 2. If we choose to dualize on node 4 (which we have seen 
is also 
strongly coupled), the resulting quiver (and the superpotential) is not of the 
same form as the one that we started with. It is straightforward to see this by Seiberg
dualizing on node 4, the gauge groups become 
\begin{eqnarray}
SU(N)_1\times SU(N+M)_3\times SU(2N+M)_4\times 
SU(N+3M)_2,
\end{eqnarray}
which is inconsistent with the original structure of the 
quiver. So the choice of the node is 
crucial for the cascade to work.

%---------------------------------------------------------

\section{\bf The Cascade Step from the $L^{a,b,c}$ Geometry} \label{calcsec}

The goal of this section is to demonstrate that, as in the conifold case, 
for an arbitrary 
$L^{a,b,c}$ with
 co-prime\footnote{For non co-prime
$(a,b,c,d)$ the corresponding $L^{a,b,c}$ space is singular and degrees of 
freedom at the singularity may become important, and so it no longer 
suffices to consider the topology alone. In that case, perhaps the 
equivariant homology 
might determine the cascade structure.} $a$, $b$, $c$ and $d\equiv a+b-c$
there is a 3-cycle $\Sigma$ satisfying (\ref{eq:g3}) for $M=1$
(and therefore for any $M$):

\begin{equation} \label{eq:g3x}
\int_{\Sigma} F_3 = 1.
\end{equation}
We will pursue the following strategy.  A Calabi-Yau cone over $L^{a,b,c}$  is actually
a K\"ahler quotient $\mathbb{C}^4/\!\!/U(1)$, namely a gauged linear $\sigma$-model (GLSM) with
$U(1)$ charges $(a,b,-c,-d)$ \cite{Martelli:2004wu}. Let us denote the $\mathbb{C}^4$
coordinates by $z_i$ with $i=1, \ldots,4$. For each $i$ there is a 
3-submanifold $\Sigma_i$ in $L^{a,b,c}$  defined by $z_i=0$. These  3-cycles are calibrated 
and therefore supersymmetric, while their volumes correspond to the $R$-charges of various fields 
in the dual gauge theory\footnote{See \cite{KMS} for the relation between the 3-cycles and 
mesonic operators in the gauge theory.}.
We will explicitly show that for an arbitrary $L^{a,b,c}$ there are two 3-cycles
$\Sigma_{3}$ and $\Sigma_{4}$ satisfying:

\begin{equation}
\label{eq:cd}
\int_{\Sigma_{3}} F_{3} = c 
\qquad
\textrm{and}
\qquad
\int_{\Sigma_{4}} F_{3} = d.
\end{equation}
Since $c$ and $d$ are co-prime the Euclidean equation $n c+m d=1$  always has a solution.
Finally, using the integers $m$ and $n$ we can construct a linear combination 
of $\Sigma_{3}$ and $\Sigma_{4}$ satisfying 
(\ref{eq:g3x}).
As we have already mentioned, for $c=d$ the $L^{a,b,c}$  geometry reduces to 
$Y^{p,q}$ and we refer the reader to \cite{HEK} for the detailed calculation in this case.
We will briefly address this case at the end of the section.

The $L^{a,b,c}$  geometry can be briefly summarised as follows.
The Sasaki-Einstein metric is given by:

\begin{equation}  
\label{eq:SE}
ds_5^2 = ds_4^2 + (d \psi^\p + A)^2,
\end{equation}
where the 4-dimensional metric is\footnote{Here we adopt the notation of 
\cite{Martelli:2005wy}.}:

\begin{equation}
ds_4^2 = \frac{(\eta-\xi)}{2 F(\xi)} d \xi^2 + \frac{2 F(\xi)}{(\eta-\xi)} 
(d \Phi+ \eta d\Psi)^2
          + \frac{(\eta-\xi)}{2 G(\eta)} d \eta^2 + \frac{2 
G(\eta)}{(\eta-\xi)} (d \Phi+ \xi d \Psi)^2,
\end{equation}
with

\begin{equation}
\label{eq:FG}
F(\xi) = 2 \xi (\alpha-\xi) (\alpha-\beta-\xi)
\qquad
\textrm{and}
\qquad
G(\eta) = - 2 \eta (\alpha-\eta) (\alpha-\beta-\eta) -2 
\end{equation}
for constant $\alpha$ and $\beta$, and the 1-form $A$ is:

\begin{equation}
A=-\half \left(  (\eta +\xi) d \Phi + \eta \xi d \Psi\right).
\end{equation}
The coordinates $\eta$ and $\xi$ vary between two adjacent roots of the 
polynomials $F(\xi)$ and $G(\eta)$ respectively. In particular, 
$0 \le \xi \le \alpha-\beta$.
The angular coordinates $\Phi$ and $\Psi$ are defined by:

\begin{equation}
\Phi \equiv \frac{\psi}{2 \beta}
\qquad
\textrm{and}
\qquad 
\Psi \equiv \frac{1}{\alpha-\beta} \left( \frac{\phi}{2 \alpha} - \frac{\psi}{2 
\beta} \right),
\end{equation}
where both $\phi$ and $\psi$ are $2\pi$-periodic. The regularity of the entire 5-dimensional metric
imposes a complicated relation between the constants $\alpha$ and $\beta$.
 
Now let us address the RR 3-form $F_3$. For the 10$d$ solution to be 
supersymmetric, 
the form $\Omega\equiv H_3 - i F_3$  has to be $(2,1)$ \cite{Grana:2000jj,Grana:2001xn}. Moreover, 
on a Calabi-Yau cone over a Sasaki-Einstein space, the $(2,1)$-form is
necessarily of the form:

\begin{equation}
\Omega_{(2,1)} = K \left( \frac{dr}{r} + i (d \psi^\p + A) \right) \wedge \omega_{(1,1)},
\end{equation}
where $K$ is a constant and
$\omega_{(1,1)} $ is  a $(1,1)$ K\"ahler form on the 4-dimensional base of
the 5-dimensional SE metric (\ref{eq:SE}). For $L^{a,b,c}$
it is:

\begin{equation}
\omega_{(1,1)}  = \frac{1}{(\eta-\xi)^2} \left( d(\eta -\xi) \wedge d \Phi + ( \eta 
d \xi -  \xi d \eta) \wedge d\Psi \right)  .
\end{equation}
We will be interested in the 3-cycles $\Sigma_3$ and $\Sigma_4$, which correspond to 
$(\xi=0, \phi=\textrm{const})$ and $(\xi=\alpha-\beta, \psi=\textrm{const})$
respectively.
The integration over these cycles yields:

\begin{equation}
\int_{\Sigma_{3}} F_{3} = K \frac{\pi}{\beta} \left( \frac{1}{\eta_2} - \frac{1}{\eta_1}\right) \Delta \psi^\p  
\quad
\textrm{and}
\quad
\int_{\Sigma_{4}} F_{3} = K \frac{\pi}{\alpha} \left( \frac{1}{\eta_2-(\alpha-\beta)} 
                                        - \frac{1}{\eta_1-(\alpha-\beta)}\right) \Delta \psi^\p  ,
\end{equation}
where $\Delta \psi^\p$ is the period of $\psi^\p$ and
$\eta_{1,2}$ are the two adjacent roots of $G(\eta)$.
Remarkably, these roots are related to the parameters $\alpha$ and $\beta$
by
\footnote{In deriving this formula, we used (3.33) of \cite{Franco:2005sm} 
with $x_i=\alpha-\eta_i$ and the explicit form
of $G(\eta)$ in (\ref{eq:FG}).}:

\begin{equation}
\frac{\alpha (\eta_2-(\alpha-\beta))  (\eta_1-(\alpha-\beta)) }
            {\beta  \eta_2 \eta_1   } = \frac{c}{d} .
\end{equation}
Thus setting 

\begin{equation}
K = \frac{\beta}{\pi  \Delta \psi^\p} \frac{\eta_1 \eta_2}{\eta_2-\eta_1} c
\end{equation}
we arrive at (\ref{eq:cd}), which in turn leads to (\ref{eq:g3x}) as we have already explained above.
Remarkably, we could have considered the cycles $\Sigma_1$ and $\Sigma_2$ 
located at $\eta=\eta_1$
and $\eta=\eta_2$ respectively. Similarly, with a proper choice of the 
constant $K$
this  yields:

\begin{equation}
\int_{\Sigma_{1}} F_{3} = a 
\qquad
\textrm{and}
\qquad
\int_{\Sigma_{2}} F_{3} = b
\end{equation}
and again, the Euclidean equation $n a+m b=1$  always has a solution since $a$ and $b$ are co-prime.
Finally, let us briefly review the $c=d$ case. In other words
we have a $Y^{p,q}$ space with $p\equiv c$ and $q\equiv c-a=b-c$.
Since the $U(1)$ factor in the isometry group is now enlarged to $SU(2)$
there are only three independent 3-cycles $\Sigma_{1}$, $\Sigma_{2}$ and 
$\Sigma_{3}$. 
These supersymmetric 3-cycles where investigated in \cite{HEK}.
It was found that for a certain value of the normalisation constant one obtains:
\begin{equation}
\int_{\Sigma_{1}} F_{3} = p-q 
\qquad
\int_{\Sigma_{2}} F_{3} = p+q
\qquad
\textrm{and}
\qquad
\int_{\Sigma_{3}} F_{3} = p,
\end{equation}
which just reproduces our results for $c=d$.
Furthermore, since $p$ and $p-q$  (alternatively $p$ and $p+q$) are co-prime
we can use $\Sigma_{1}$ and $\Sigma_{3}$ to construct the 3-cycle 
$\Sigma$
satisfying (\ref{eq:g3x}). 
This completes the proof of the main claim of the paper.

%---------------------------------------------------------

\section{Conclusions}
\label{conc}
$L^{a,b,c}$, at least when $a,\ b,\ c$\ and $d=a+b-c$ are relatively prime, 
is non-singular and diffeomorphic to $S^2\times S^3$.  
In this note we have argued that this fact, together with RR charge conservation, 
is sufficient to restrict the form of possible cascades.  We considered cascades 
in which each duality corresponds to an NS5-brane that sweeps out the 3-sphere, 
and argued that the 3-form RR flux on the 3-sphere implies that such a process necessarily destroys 
a number of D3-branes equal to the number of D5-branes, corresponding to a Seiberg duality in the gauge theory.  
We also checked that for any number of D5-branes there exists a 3-form representing the corresponding 
de Rham cohomology class which is $(2,1)$, as is required by supersymmetry.

This result applies more generally.  Only the integrals of the various 
forms over the cycles were important, and so it suffices to consider an 
integral sublattice of the de Rham cohomology, which in this case is 
isomorphic to the integral cohomology.\footnote{In general the integral 
cohomology may also contain torsion subgroups, which may lead to 
interesting variations of the dual gauge theories corresponding to 
discrete torsion fluxes in the string theory compactification.}  
In particular, cascades caused by 5-branes sweeping out 3-cycles appear 
to never change the form of the quiver, because the 5-branes violate 
D3-brane charge which is classified by the zeroth cohomology of the 
compact space, which is always one-dimensional as the space is connected.  
Thus each step in the cascade corresponds to a change in a single 
parameter.  If there are multiple 3-cycles, then the minimal cascade is 
simply the greatest common divisor of the number of D3-branes created by 
5-branes wrapping the various 3-cycles.  Exotic cascades may be possible 
if one also considers processes in which D7-branes nucleate, for example 
a D7-brane sweeping out a 5-cycle supporting a nontrivial $H$-flux will 
violate the D5-brane charge wrapping the 2-cycle dual to the $H$-flux in 
the 5-cycle.  In practice many of these examples remain out of reach as 
they require an understanding of S-duality in the presence of D7-branes.

The self-similarity of these cascades appears to be in contradiction with the duality walls that are 
predicted from a purely gauge-theoretic point of view.  It may be that this supergravity analysis is too naive, 
that one must consider also the physics at the tip of the cone, where many different cycles exists and may come 
in and out of existence via geometric transitions, however branes wrapping such cycles tend to lead to chiral 
anomalies in the gauge theory.  Another possibility is that D7-brane processes must be considered in such cases.  
However, it may also be that in the gauge theory analysis, which relies on an analogy with a theory with 
a single simple gauge group, approximating the others to be global symmetries in the IR despite the sign of 
their beta functions, is invalid.

%---------------------------------------------------------

%\appendix
%\section{}

%\label{App}

%===============
\section* {Acknowledgements}
%===============

It is our pleasure to thank I.~Klebanov for suggesting this problem and
R.~Argurio for fruitful discussions and a careful reading of the manuscript. 
We are also honoured to thank F.~Bigazzi  and D.~Persson for invaluable comments.
This work is partially supported by IISN-Belgium (convention 4.4505.86), by the ``Interuniversity Attraction 
Poles Programme -- Belgian Science Policy" and by the European 
Commission programme MRTN-CT-2004-005104, in which the authors are 
associated to V.U. Brussel.

%\bibliography{EKK}

\begin{thebibliography}{--} 


\bibitem{KW}
I. R. Klebanov and E. Witten,
{\it Superconformal Field Theory on Threebranes at a Calabi-Yau Singularity},
[{\tt arXiv:hep-th/9807080}].

\bibitem{KN}
I. R. Klebanov and N. A. Nekrasov,
{\it Gravity duals of fractional branes and logarithmic RG flow},
[{\tt arXiv:hep-th/9911096}].

\bibitem{KT}
I. R. Klebanov and A. A. Tseytlin,
{\it Gravity duals of supersymmetric SU(N) x SU(N+M) gauge theories},
[{\tt arXiv:hep-th/0002159}].

\bibitem{KS}
I. R. Klebanov and M. J. Strassler,
{\it Supergravity and a Confining Gauge Theory: Duality Cascades and $\chi$SB-Resolution of Naked Singularities},
[{\tt arXiv:hep-th/0007191}].

\bibitem{Gauntlett:2004yd}
J.~P. Gauntlett, D.~Martelli, J. Sparks and D. Waldram, 
{\it {S}asaki-{E}instein metrics on ${S}^2 \times {S}^3$}, 
[{\tt arXiv:hep-th/0403002}].  

\bibitem{Bertolini:2004xf}
M.~Bertolini, F.~Bigazzi and A.~L. Cotrone, {\it New checks and subtleties for
  ads/cft and a-maximization},  
[{\tt arXiv:hep-th/0411249}].  

\bibitem{Benvenuti:2004dy}
S.~Benvenuti, S.~Franco, A.~Hanany, D.~Martelli and J.~Sparks, {\it An
  infinite family of superconformal quiver gauge theories with
  {S}asaki-{E}instein duals}, 
[{\tt arXiv:hep-th/0411264}].  

\bibitem{Cvetic:2005ft}
M.~Cvetic, H.~Lu, D.~N. Page and C.~N. Pope, {\it New {E}instein-{S}asaki
  spaces in five and higher dimensions}, 
[{\tt  arXiv:hep-th/0504225}].  

\bibitem{Benvenuti:2005cz}
S.~Benvenuti and M.~Kruczenski, {\it Semiclassical strings in
  {S}asaki-{E}instein manifolds and long operators in {N} = 1 gauge theories},
[{\tt arXiv:hep-th/0505046}].  

\bibitem{Benvenuti:2005ja}
S.~Benvenuti and M.~Kruczenski, {\it From {S}asaki-{E}instein spaces to quivers
  via {BPS} geodesics: ${L}^{p,q,r}$},
[{\tt arXiv:hep-th/0505206}].  

\bibitem{Franco:2005sm}
S.~Franco, A.~Hanany, D.~Martelli, J.~Sparks, D.~Vegh and B.~Wecht, 
{\it Gauge theories from toric geometry and brane
  tilings}, 
[{\tt arXiv:hep-th/0505211}].  

\bibitem{Cvetic:2005vk}
M.~Cvetic, H.~Lu, D.~N. Page and C.~N. Pope, {\it New {E}instein-{S}asaki and
  {E}instein spaces from {K}err-de {S}itter},
[{\tt arXiv:hep-th/0505223}].  

\bibitem{Butti:2005sw}
A.~Butti, D.~Forcella and A.~Zaffaroni, {\it The dual superconformal theory
  for ${L}^{p,q,r}$ manifolds},  
[{\tt arXiv:hep-th/0505220}].  

\bibitem{Martelli:2005wy}
  D.~Martelli and J.~Sparks, {\it Toric {S}asaki-{E}instein metrics on ${S}^2
  \times {S}^3$},  
[{\tt arXiv:hep-th/0505027}].  

\bibitem{Berenstein:2005xa}
  D.~Berenstein, C.~P.~Herzog, P.~Ouyang and S.~Pinansky,
  {\it Supersymmetry breaking from a Calabi-Yau singularity},
[{\tt arXiv:hep-th/0505029}]. 

\bibitem{Franco:2005zu}
  S.~Franco, A.~Hanany, F.~Saad and A.~M.~Uranga,
   {\it Fractional branes and dynamical supersymmetry breaking},
[{\tt arXiv:hep-th/0505040}].  

\bibitem{Bertolini:2005di}
  M.~Bertolini, F.~Bigazzi and A.~L.~Cotrone,
   {\it Supersymmetry breaking at the end of a cascade of Seiberg dualities},
[{\tt arXiv:hep-th/0505055}].

\bibitem{Brini:2006ej}
  A.~Brini and D.~Forcella,
{\it Comments on the non-conformal gauge theories dual to Y(p,q) manifolds},
[{\tt arXiv:hep-th/0603245}].

\bibitem{Fiol:2002ah}
B. Fiol,
{\it Duality cascades and duality walls},
[{\tt arXiv:hep-th/0205155}].


\bibitem{Franco:2003ja}
S. Franco, A. Hanany, Y.-H. He and P. Kazakopoulos, 
{\it Duality walls, duality trees and fractional branes},
[{\tt arXiv:hep-th/0306092}].

\bibitem{Franco:2004jz}
S. Franco, Y.-H. He, C. Herzog and J. Walcher, 
{\it Chaotic duality in string theory},
[{\tt arXiv:hep-th/0402120}].

\bibitem{CdO}
P. Candelas and X. C. de la Ossa,
{\it Comments on Conifolds},
Nucl.\ Phys.\ B {\bf 342}, 246-268 (1990).

\bibitem{EK}
J. Evslin and S. Kuperstein, 
{\it Trivializing and orbifolding the conifold's base},
[{\tt arXiv:hep-th/0702041}].

\bibitem{KPV}
S. Kachru, J. Pearson and H. L. Verlinde,
{\it Brane/Flux Annihilation and the String Dual of a Non-Supersymmetric Field Theory},
[{\tt arXiv:hep-th/0112197}].

\bibitem{MeCascade}
J. Evslin,
{\it The Cascade is a MMS Instanton},
[{\tt arXiv:hep-th/0405210}].

\bibitem{Evslin:2007ti}
  J.~Evslin and L.~Martucci,
  {\it D-brane networks in flux vacua, generalized cycles and calibrations},
[{\tt arXiv:hep-th/0703129}].  

\bibitem{MMS}
J. M. Maldacena, G. W. Moore and N. Seiberg, 
{\it D-brane instantons and K-theory charges},
[{\tt arXiv:hep-th/0108100}].

\bibitem{BDS}
C. Bachas, M. R. Douglas and C. Schweigert,  
{\it Flux stabilization of D-branes},
[{\tt arXiv:hep-th/0003037}].

\bibitem{APS}
  P.~C.~Argyres, M.~R.~Plesser and N.~Seiberg,
  {\it The Moduli Space of N=2 SUSY {QCD} and Duality in N=1 SUSY {QCD}},
  [{\tt arXiv:hep-th/9603042}].

\bibitem{Hanany:2005ve}
A. Hanany and K. D. Kennaway,
{\it Dimer models and toric diagrams},
[{\tt arXiv:hep-th/0503149}].  

\bibitem{Grana:2000jj}
M. Grana and J. Polchinski, 
{\it Supersymmetric three-form flux perturbations on AdS(5)},
[{\tt arXiv:hep-th/0009211}].

\bibitem{Grana:2001xn}
M. Grana and J. Polchinski, 
{\it Gauge / gravity duals with holomorphic dilaton},
[{\tt arXiv:hep-th/0106014}].

\bibitem{Martelli:2004wu}
D.~Martelli and J.~Sparks, 
{\it Toric geometry, {S}asaki-{E}instein manifolds and a new infinite class of {AdS/CFT} duals}, 
[{\tt arXiv:hep-th/0411238}].  

\bibitem{Feng:2000mi}
B. Feng, A. Hanany and Y.-H. He,
{\it D-brane gauge theories from toric singularities and toric duality},
[{\tt arXiv:hep-th/0003085}].

\bibitem{HEK}
C. P. Herzog, Q. J. Ejaz and I. R. Klebanov, 
{\it Cascading RG flows from new Sasaki-Einstein manifolds}, 
[{\tt arXiv:hep-th/0412193}].  

\bibitem{Burrington:2005zd}
B. A. Burrington, J. T.  Liu, M. Mahato and L. A. Pando-Zayas, 
{\it Towards supergravity duals of chiral symmetry breaking in
                  Sasaki-Einstein cascading quiver theories}, 
[{\tt  arXiv:hep-th/0504225}].  

\bibitem{Gepner:2005zt}
D. Gepner and S. S. Pal,  {\it Branes in L(p,q,r)}, 
[{\tt  arXiv:hep-th/0505039}].  


\bibitem{kreview}
J. Evslin, {\it What does(n't) K-theory classify?},
  [{\tt arXiv:hep-th/0610328}].

\bibitem{KMS}
S.~Kuperstein, O.~Mintkevich and J.~Sonnenschein,
{\it On the pp-Wave Limit and the BMN Structure of New Sasaki-Einstein Spaces},
[{\tt arXiv:hep-th/0609194}].













\end{thebibliography}

%\newpage
\bibliographystyle{unsrt}

\end{document}